\documentclass[aps,prl,twocolumn,superscriptaddress,amsfont,graphicx,nofootinbib,preprintnumbers]{revtex4}%
\usepackage{color,graphicx,epsfig}
\usepackage{ifpdf}
\usepackage{amsmath}
\usepackage{bm}
\usepackage{color}
\usepackage[english]{babel}
\usepackage{graphicx}%
\usepackage{amsfonts}%
\usepackage{amssymb}
\usepackage{braket}
\usepackage{hyperref}
\usepackage{enumerate}

\definecolor{nicered}{rgb}{0.7,0.1,0.1}
\definecolor{nicegreen}{rgb}{0.1,0.5,0.1}
\hypersetup{colorlinks,citecolor= nicegreen,linkcolor= nicered}

\newcommand{\beq}{\begin{equation}}
\newcommand{\eeq}{\end{equation}}
\newcommand{\bea}{\begin{eqnarray}}
\newcommand{\eea}{\end{eqnarray}}

\definecolor{Red}{rgb}{1.,0.,0.}

\def\taun{{\cal T}_N}

\def\tauncut{{\cal T}_N^{cut}}
\def\tauonecut{{\cal T}_1^{cut}}

\def\mysection#1{{{\bf #1}.~}}
\def\OMIT#1{}

\begin{document}

\def\Maryland{Maryland Center for Fundamental Physics, University of Maryland, College Park, Maryland 20742, USA}
\def\Argonne{High Energy Physics Division, Argonne National Laboratory, Argonne, IL 60439, USA}
\def\Northwestern{Department of Physics \& Astronomy, Northwestern University, Evanston, IL 60208, USA}
\def\Fermilab{Fermilab, P.O.Box 500, Batavia, IL 60510, USA}
\def\Durham{Institute for Particle Physics Phenomenology, Department of Physics, University of Durham, Durham, DH1 3LE, UK}

\preprint{FERMILAB-PUB-15-519-T}

\title{$Z$-boson production in association with a jet at next-to-next-to-leading order in perturbative QCD}

\author{Radja Boughezal}     
\email[Electronic address:]{rboughezal@anl.gov}
\affiliation{\Argonne}

\author{John Campbell}     
\email[Electronic address:]{johnmc@fnal.gov}
\affiliation{\Fermilab}

\author{R. Keith Ellis}     
\email[Electronic address:]{keith.ellis@durham.ac.uk}
\affiliation{\Durham}

\author{Christfried Focke}
\email[Electronic address:]{christfried.focke@northwestern.edu}
\affiliation{\Northwestern}

\author{Walter Giele}     
\email[Electronic address:]{giele@fnal.gov}
\affiliation{\Fermilab}

\author{Xiaohui Liu}
\email[Electronic address:]{xhliu@umd.edu}
\affiliation{\Maryland}

\author{Frank Petriello}     
\email[Electronic address:]{f-petriello@northwestern.edu}
\affiliation{\Argonne}
\affiliation{\Northwestern}

\date{\today}
\begin{abstract}
We present the first complete calculation of $Z$-boson production in association with a jet in hadronic collisions through next-to-next-to-leading order in perturbative QCD. Our computation uses the recently-proposed $N$-jettiness subtraction scheme to regulate the infrared divergences that appear in the real-emission contributions.  We present phenomenological results for 13 TeV proton-proton collisions with fully realistic fiducial cuts on the final-state particles.  The remaining theoretical uncertainties after the inclusion of our calculations are at the percent-level, making the $Z$+jet channel ready for precision studies at the LHC Run II.
\end{abstract}

\maketitle

\section{Introduction} \label{sec:intro}

The production of a $Z$-boson in association with a jet is an important process for the physics program of the Large Hadron Collider (LHC).  It serves as a background to searches for supersymmetry and for dark matter in the mono-jet channel, and in measurements of properties of the Higgs boson.  The measurement of the $Z$+jet process can also be used to improve the determination of the gluon distribution function.  For all of these purposes a precision Standard Model prediction of this process is highly desirable. 

The next-to-leading order (NLO) corrections in the strong coupling constant for $Z$+jet production have been known for some time~\cite{Giele:1993dj}.  The NLO electroweak corrections were considered in Ref.~\cite{EWcor}.  First partial results for next-to-next-to-leading order (NNLO) corrections to the $qg$, $q\bar{q}$, and $gg$ partonic channels in the leading-color approximation were recently presented~\cite{Ridder:2015dxa}.  However, none of these results are suitable for precision phenomenology at the LHC.  The scale uncertainty of the NLO calculation is comparable to the combination of all other experimental systematic errors at high transverse momenta of the leading jet~\cite{Aad:2013ysa,Khachatryan:2014zya}.  Inclusion of partial NNLO corrections is a first step to improve this situation, but even small partonic channels can shift the distribution shapes in non-trivial ways, in particular at high transverse momentum~\cite{radcortalk}.  A complete calculation is highly desirable.

In this paper we report on a complete calculation of $Z$-boson production in association with a jet at NNLO in perturbative QCD, including all partonic channels and maintaining the full color dependence.  We investigate the effects of higher-order QCD corrections on the kinematics of the $Z$-boson, the leading jet, and the leptons arising from the $Z$-boson decay in 13 TeV LHC collisions. Fully realistic acceptance cuts are imposed on the final-state particles.  We find that the NNLO corrections are at the percent-level over most of the studied phase space, and have minimal kinematic dependence.  After our calculation the $Z$+jet channel is ready for a precision comparison with the upcoming data from the LHC Run II.

To derive these predictions we use the recently-proposed $N$-jettiness subtraction technique~\cite{Boughezal:2015dva,Gaunt:2015pea}, which has been used to provide the first complete predictions for both $W$-boson and Higgs boson production in association with a jet at NNLO~\cite{Boughezal:2015dva,Boughezal:2015aha}.\footnote{The Higgs plus jet process has also been calculated using other NNLO subtraction methods~\cite{Boughezal:2013uia,Chen:2014gva,Boughezal:2015dra}.}  We incorporate our results into a new version of the MCFM program~\cite{Campbell:2010ff} that supports NNLO calculations using the $N$-jettiness framework.  An interesting feature of the computational algorithm used in this new version is that it exhibits strong scaling to many thousands of nodes, and can run on modern supercomputing platforms.  This makes possible calculations and phenomenological studies that were previously intractable.  We will discuss the details of our approach for color-singlet production in a recent publication~\cite{wip1}.
 
\section{Theoretical Framework} \label{sec:theory}

We sketch here the $N$-jettiness subtraction scheme.  The implementation of this scheme used in obtaining our results was presented in Ref.~\cite{Boughezal:2015dva}.  Another description of the method is also given in Ref.~\cite{Gaunt:2015pea}.  

We begin with the definition of the $N$-jettiness variable $\taun$, a global event shape designed to veto final-state jets~\cite{Stewart:2010tn}:
\begin{equation}
\label{eq:taudef}
{\cal T}_N = \sum_k \text{min}_i \left\{ \frac{2 p_i \cdot q_k}{Q_i}\right\}.
\end{equation}
The subscript $N$ denotes the number of jets desired in the final state; for the $Z$+jet process considered here, $N=1$.  Values of ${\cal T}_1$ near zero indicate a final state containing a single narrow energy deposition, while larger values denote a final state containing two or more well-separated energy depositions.   The $p_i$ are light-like reference vectors for each of the initial beams and final-state jets in the problem.  The reference vectors for the final-state jets can be determined by using a jet algorithm, as discussed in Refs.~\cite{Stewart:2010tn,Jouttenus:2011wh}.  The determination of the $p_i$ is insensitive to the choice of jet algorithm in the small-$\tauncut$ limit~\cite{Stewart:2010tn}.  The $q_k$ denote the four-momenta of any final-state radiation.  The $Q_i$ characterize the hardness of the beam-jets and final-state jets.  We set $Q_i = 2 E_i$, twice the lab-frame energy of each jet.  

We briefly outline the procedure through which we use $\taun$ to obtain the complete NNLO correction to the $Z$+jet process.  The NNLO cross section consists of contributions with Born-level kinematics, and processes with either one or two additional partons radiated.  We partition the phase space for each of these terms into regions above and below a cutoff on $\taun$, which we label $\tauncut$:
\begin{equation}
\label{eq:partition}
\begin{split}
\sigma_{NNLO} &= \int {\rm d}\Phi_N \, |{\cal M}_{N}|^2 +\int {\rm d}\Phi_{N+1} \, |{\cal M}_{N+1}|^2 \, \theta_N^{<} \\
&+\int {\rm d}\Phi_{N+2} \, |{\cal M}_{N+2}|^2 \, \theta_N^{<}+\int {\rm d}\Phi_{N+1} \, |{\cal M}_{N+1}|^2 \, \theta_N^{>} \\
&+\int {\rm d}\Phi_{N+2} \, |{\cal M}_{N+2}|^2 \, \theta_N^{>} \\
& \equiv  \sigma_{NNLO}(\taun < \tauncut)+\sigma_{NNLO}(\taun > \tauncut).
\end{split}
\end{equation}
We have abbreviated $\theta_N^{<} = \theta(\tauncut-\taun)$ and $\theta_N^{>} = \theta(\taun-\tauncut)$.  The first three terms in this expression all have $\taun<\tauncut$, and have been collectively denoted as $\sigma_{NNLO}(\taun < \tauncut)$.  The remaining two terms have $\taun>\tauncut$, and have been collectively denoted as $\sigma_{NNLO}(\taun > \tauncut)$.  Contributions with Born-level kinematics necessarily have $\taun=0$.  Similar partitionings of phase space have proven useful in the context of merging fixed-order calculations with parton showers in the effective-theory framework~\cite{Alioli:2012fc}.

The critical point that allows us to compute the cross section to NNLO below $\tauncut$ is the existence of a factorization theorem that gives an all-orders description of $N$-jettiness for small $\taun$~\cite{Stewart:2009yx,Stewart:2010pd}.  
Using this result, the cross section with $\taun$ less than some value $\tauncut$ can be written in the schematic form
\begin{equation} \label{eq:fact}
 \sigma(\taun < \tauncut)=  \int H \otimes B \otimes B \otimes S \otimes   \left[ \prod_{n}^{N} J_n \right] +\cdots .
\end{equation}
$H$ is the hard function which encodes the virtual corrections to the process.  $B$ is the beam function, which describes the effect of radiation collinear to one of the two initial beam directions.  The general importance of the beam function in describing hadronic collisions was first realized in Ref.~\cite{Stewart:2009yx}.  It can be decomposed as a perturbative matching coefficient convolved with the usual parton distribution function.  $S$ describes the soft radiation, and $J_n$ contains the radiation collinear to a final-state jet.  
Depending on the observable and process under consideration, only a subset of these terms may be present.  
The ellipsis denotes power-suppressed terms which become negligible for $\taun \ll Q_i$. The derivation of this all-orders expression in the small-$\taun$ limit uses the machinery of Soft-Collinear Effective Theory~\cite{Bauer:2000ew}.  Upon expansion to fixed-order in the strong coupling constant, Eq.~(\ref{eq:fact}) reproduces the fixed-order cross section $\sigma_{NNLO}(\taun < \tauncut)$ for low $\tauncut$ needed in Eq.~(\ref{eq:partition}).  The two-loop virtual corrections needed for the NNLO hard function are known for the $Z$+jet process~\cite{Gehrmann:2011ab,Gehrmann:2013vga}.\footnote{We clarify here an unclear point in Ref.~\cite{Gehrmann:2011ab}: to obtain the virtual corrections for all helicity amplitudes, only the spinor products should be conjugated in Eqs.~(2.24-2.25), not the $\alpha$, $\beta$ and $\gamma$ coefficients.}  The beam functions are known at NNLO~\cite{Gaunt:2014xga}, as are the jet functions~\cite{Becher:2006qw} and the soft function~\cite{Boughezal:2015eha}. 

A full NNLO calculation requires as well the high $\taun$ region above $\tauncut$.  However, a finite value of $\taun$ implies that there are actually at least $N+1$ resolved partons in the final state.  The cross section above the cut  can be obtained from a NLO calculation containing an additional jet.  We must choose $\tauncut$ much smaller than any other kinematical invariant in the problem in order to avoid power corrections to Eq.~(\ref{eq:fact}) below the cutoff.  We discuss the validation of the explicit $\tauncut$ values used in our numerical results in a later section.

\section{Validation of the Formalism} \label{sec:formalism}

We now discuss how we obtain the various components of Eq.~(\ref{eq:partition}) needed to obtain the complete cross section at NNLO.  Above $\tauncut$ we need a NLO calculation of $Z$+2-jets.  We use an improved version of  MCFM~\cite{Campbell:2010ff} optimized to handle the $N$-jettiness subtraction scheme to obtain this contribution efficiently.  Upon integration over the phase space of the final-state leptons, we can check our implementation of the hard function against PeTeR~\cite{Becher:2011fc}; we have done so and have found perfect agreement.  The calculation and validation of the necessary $N$-jettiness soft function has been detailed in a separate publication~\cite{Boughezal:2015eha}.  The necessary two-loop beam and jet functions for this process are also known~\cite{Gaunt:2014xga,Becher:2006qw}.

\begin{figure}[h]
\centering
\includegraphics[width=3.3in]{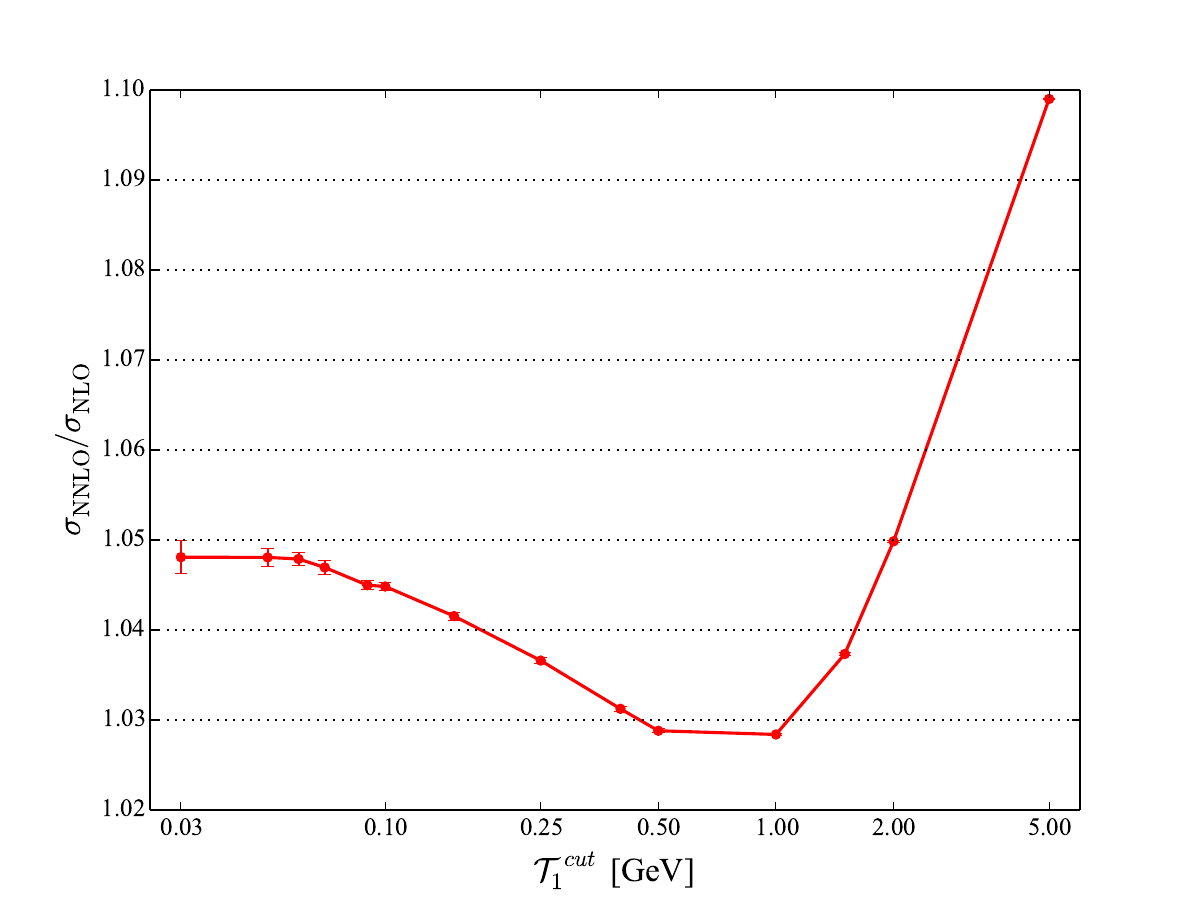}%
\caption{Plot of the NNLO cross section over the NLO result, $\sigma_{\text{NNLO}}/\sigma_{\text{NLO}}$, as a function of $\tauonecut$, for the scale choice $\mu= 2 \times \mu_0$.  The vertical bars accompanying each point indicate the integration errors.} \label{fig:taucheck}
\end{figure}

The primary check of the $N$-jettiness formalism is that the logarithmic dependence on $\tauncut$ that occurs separately in the low and high $\taun$ regions cancels when they are summed.  This requires that almost all parts of the calculation are implemented correctly and consistently; the beam, soft, and jet functions, as well as the NLO corrections to $Z$+2-jets, are probed by this check.  We show in Fig.~\ref{fig:taucheck} the results of this validation for the ratio $\sigma_{\text{NNLO}}/\sigma_{\text{NLO}}$ in 13 TeV proton-proton collisions (we note that NNLO PDFs are used in the numerator, while NLO PDFs are used in the denominator).  We have checked that the NLO cross section obtained with $N$-jettiness subtraction agrees with the result obtained with standard techniques.  These cross sections are obtained using CT14 parton distribution functions~\cite{Dulat:2015mca} at the same order in perturbation theory as the partonic cross section, and contain the following fiducial cuts on the leading final-state jet and the two leptons from CMS~\cite{Khachatryan:2014zya}: $p_T^{jet} > 30$ GeV, $|\eta_{jet}|<2.4$, $p_T^{l} > 20$ GeV, $|\eta_{l}|<2.4$ and $71\, \text{GeV} < m_{ll} < 111 \, \text{GeV}$. The ATLAS analysis is similar but with slightly different cuts~\cite{Aad:2013ysa}.  We reconstruct jets using the anti-$k_T$ algorithm~\cite{Cacciari:2008gp} with $R=0.5$.  A dynamical scale $\mu_0=\sqrt{m_{ll}^2+\sum p_{T}^{jet,2}}$ is chosen to describe this process, where the sum is over the transverse momenta of all final-state jets, and $m_{ll}$ the invariant mass of the di-lepton pair arising from the $Z$-boson decay. In this validation plot we have set the renormalization and factorization scales to $\mu_R = \mu_F = 2 \times \mu_0$; since the corrections are larger for this scale choice, it is easier to illustrate the important aspects of the $\tauonecut$ variation.

A few features can be seen in Fig.~\ref{fig:taucheck}.  First, in the region $\tauonecut < 0.08$ GeV the result becomes independent of the particular value of the cut chosen within the numerical errors.  The NNLO correction for $\mu = 2 \times \mu_0$ corresponds to an almost $+5\%$ shift in the cross section.  The plot makes clear that we have numerical control over the NNLO cross section to the per-mille level, completely sufficient for phenomenological predictions.  We observe an approximately linear dependence of $\sigma_{\text{NNLO}}$ on ${\rm ln}\left(\tauonecut\right)$ in the region $ 0.1 \, {\rm GeV} < \tauonecut < 0.5 \, {\rm GeV}$, indicating the onset of the power corrections neglected in Eq.~(\ref{eq:fact}).  These power corrections have the form $(\taun / Q) \, {\rm ln}^n(\taun / Q)$, where $n \leq 3$ at NNLO~\cite{Gaunt:2015pea} and $Q$ is a hard scale such as $p_T^{jet}$.

The other possible checks of the $N$-jettiness formalism involve comparison with other NNLO results obtained using different techniques.  We have previously checked that the agreement between Higgs+jet production as computed with $N$-jettiness and with other techniques~\cite{Boughezal:2013uia} agree at the per-mille level~\cite{Boughezal:2015aha}.  A selection of processes without final-state jets have also been computed with both $N$-jettiness subtraction and other techniques, and show a similar level of agreement~\cite{Gaunt:2015pea,wip1}.

\section{Numerical Results}

We present here numerical results for $Z$-boson production in association with a jet at NNLO.  Our central scale choice is the dynamical scale $\mu=\mu_0$, as described in the previous section.  To obtain an estimate of the theoretical errors we vary $\mu$ away from this choice by a factor of two.  We use the same cuts on the jets and leptons as described in the previous section.  We include the contributions from both the $Z$-boson and a virtual photon decaying to leptons in our numerical results.

\begin{figure}[h]
\centering
\includegraphics[width=3.3in]{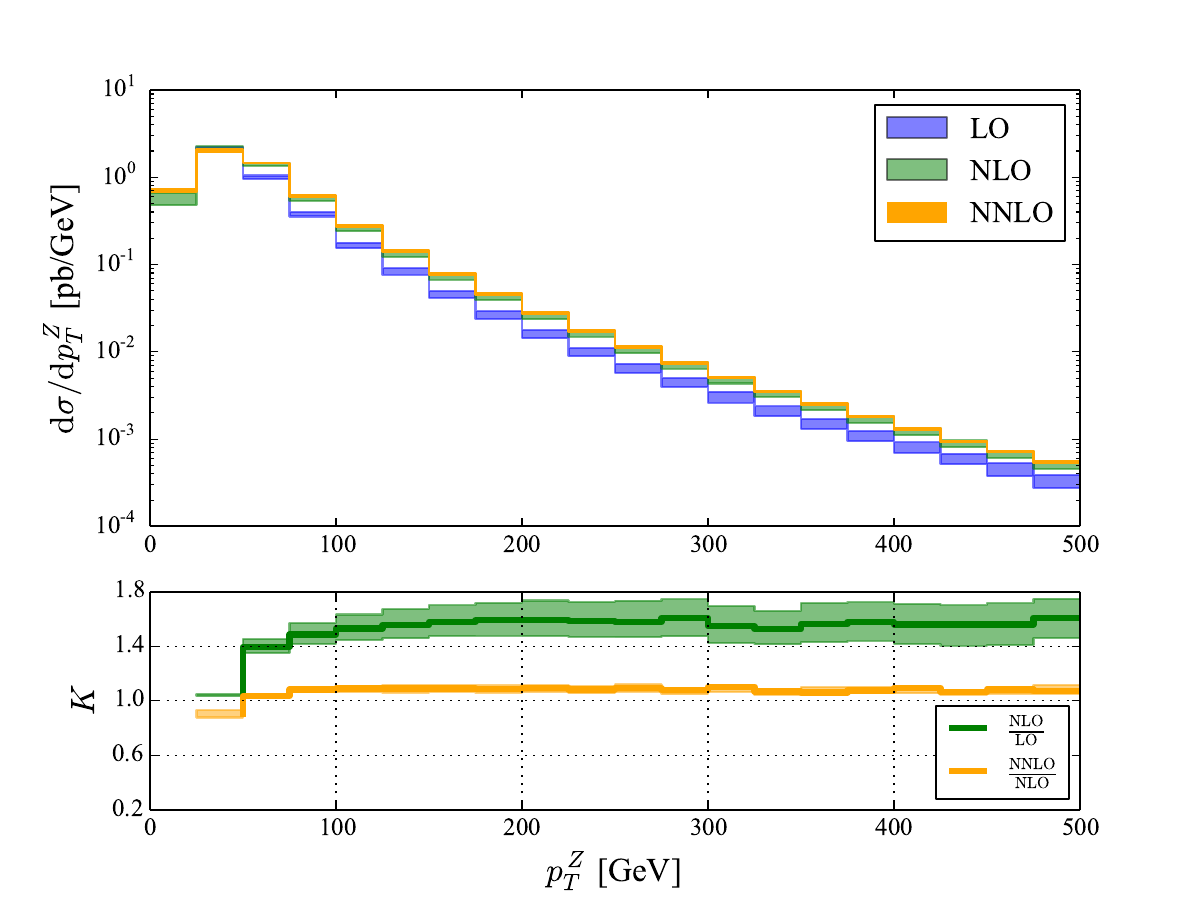}%
\caption{Plot of the $Z$-boson $p_T$ distribution at LO, NLO and NNLO in QCD perturbation theory, for 13 TeV collisions with the central scale $\mu_0=\sqrt{m_{ll}^2+\sum p_{T}^{jet,2}}$.  The $K$-factors are shown in the lower inset.} \label{fig:pTZ}
\end{figure}

\begin{figure}[h]
\centering
\includegraphics[width=3.3in]{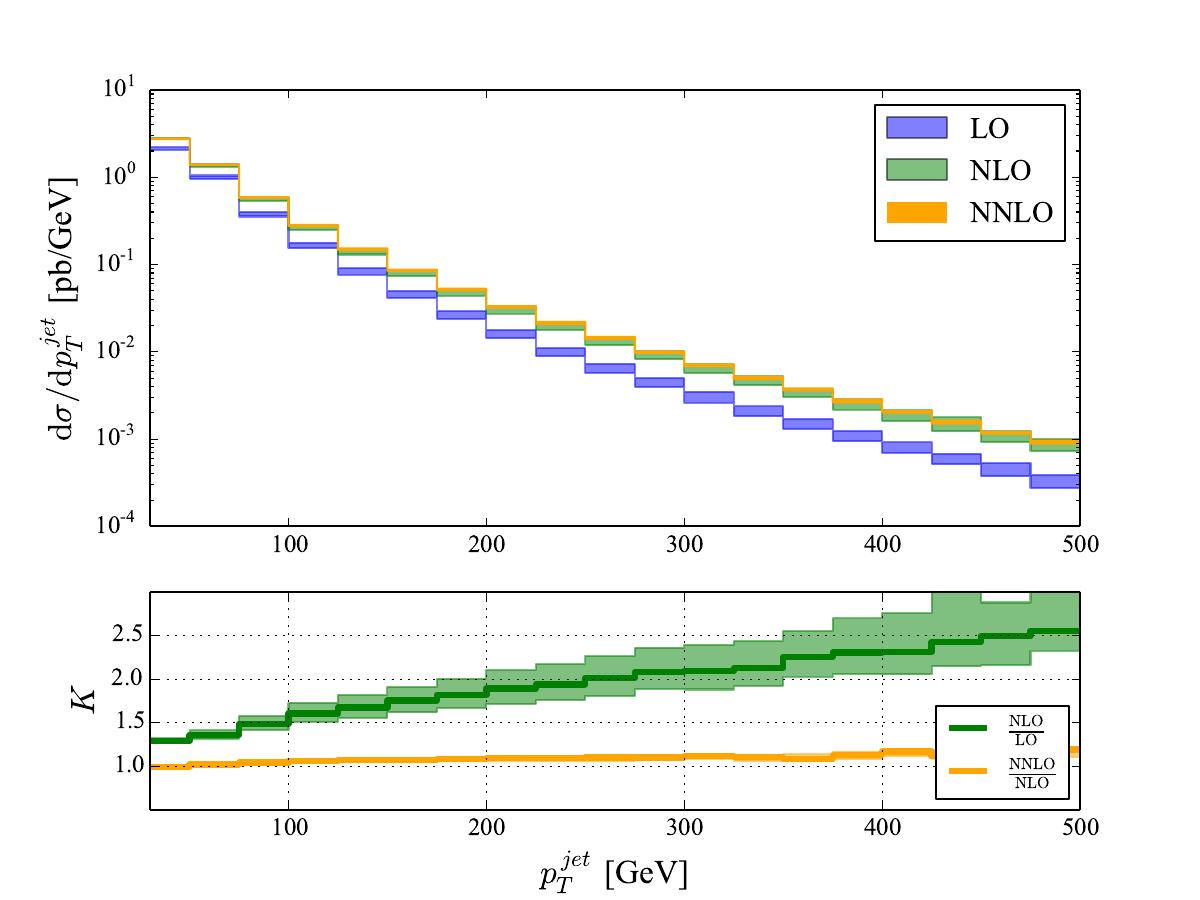}%
\caption{Plot of the leading-jet $p_T$ distribution at LO, NLO and NNLO in QCD perturbation theory, for 13 TeV collisions with the central scale $\mu_0=\sqrt{m_{ll}^2+\sum p_{T}^{jet,2}}$.  The $K$-factors are shown in the lower inset.} \label{fig:pTJ}
\end{figure}

We note that the cross sections at each order in perturbation theory for the cuts described above are:
\begin{equation}
\begin{split}
\sigma_{\text{LO}} &= 97.4^{+3.9}_{-4.4} \; \text{pb}, \\
\sigma_{\text{NLO}} &= 133.3^{+5.4}_{-4.2} \; \text{pb}, \\
\sigma_{\text{NNLO}} &= 135.6^{+0.0}_{-0.4} \; \text{pb}. \\
\end{split}
\end{equation}
The NNLO correction results in a $+1\%$ increase in the fiducial cross section.  The scale dependence is greatly reduced with respect to the NLO result.  We note that the full NNLO corrections are in good agreement with an alternative calculation found in Ref.~\cite{Ridder:2015dxa} that has recently become available. We next show the $Z$-boson transverse momentum distribution in Fig.~\ref{fig:pTZ}, focusing on the range $p_{T}^Z<500$ GeV.  The distributions at LO, NLO and NNLO in QCD perturbation theory are shown, as are the usual $K$-factors: the ratio of the NLO over the LO cross section, and the NNLO over the NLO result.  To produce this distribution and all other ones, we average the results from $\tauonecut=0.03,0.04,0.05,0.06$ GeV. A reduced scale dependence is obtained when the NNLO corrections are included, and a significantly smaller correction is observed when going from NLO to NNLO than when going from LO to NLO, indicating stability of the perturbative expansion.  A slight increase of the NNLO correction occurs as $p_{T}^Z$ is increased.  The analogous transverse momentum distribution for the leading jet is shown in Fig.~\ref{fig:pTJ}.  In this case the NLO corrections grow with $p_{T}^{jet}$, reaching a $K$-factor of 2.5 for $p_{T}^{jet}=500$ GeV.  The NNLO corrections are far more mild, but they grow with $p_{T}^{jet}$, increasing the NLO result by 10\% at $p_{T}^{jet}=500$ GeV.  It is essential to account for these corrections when comparing with measurements, as the experimental errors are only at the few-percent level in this region.

\begin{figure}[h]
\centering
\includegraphics[width=3.3in]{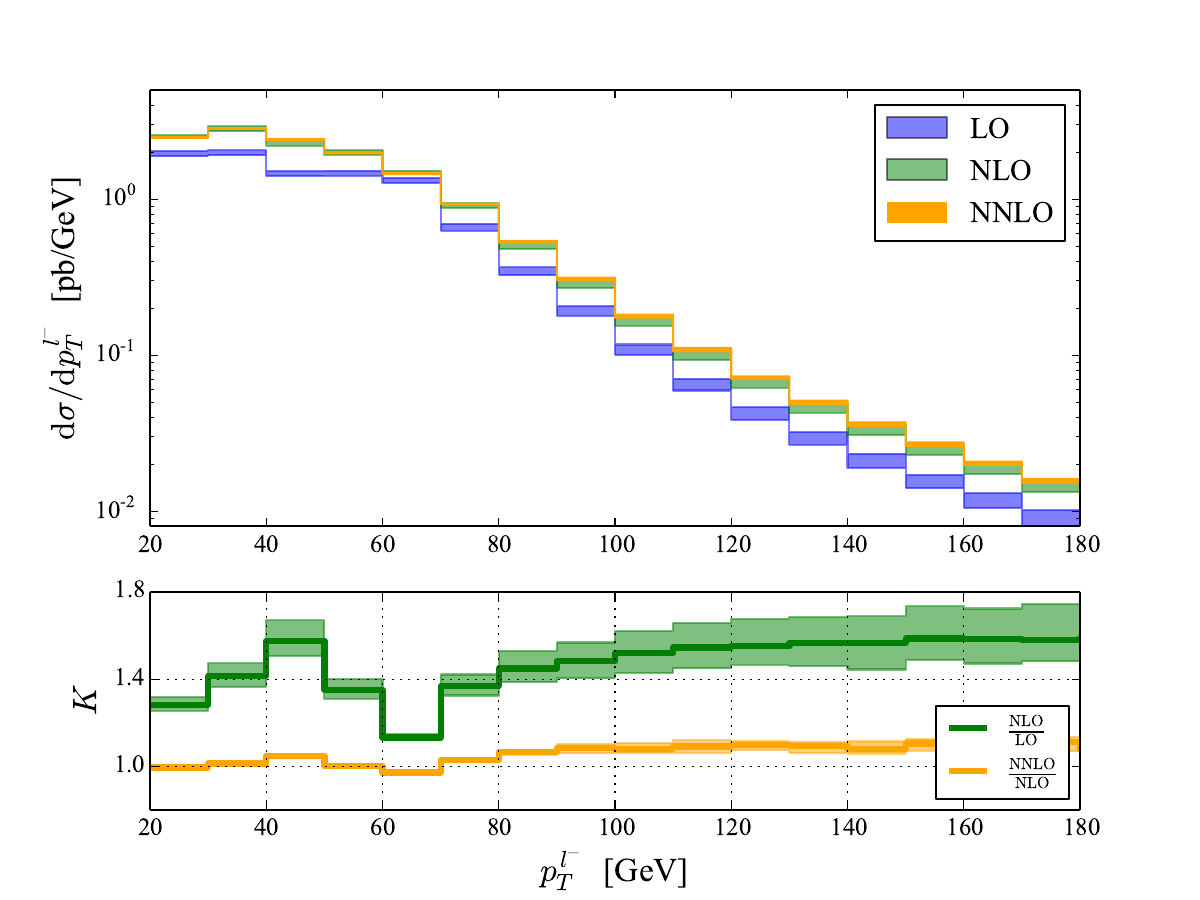}%
\caption{Plot of the lepton $p_T$ distribution at LO, NLO and NNLO in QCD perturbation theory, for 13 TeV collisions with the central scale $\mu_0=\sqrt{m_{ll}^2+\sum p_{T}^{jet,2}}$.  The $K$-factors are shown in the lower inset.} \label{fig:pTl}
\end{figure}

\begin{figure}[h]
\centering
\includegraphics[width=3.3in]{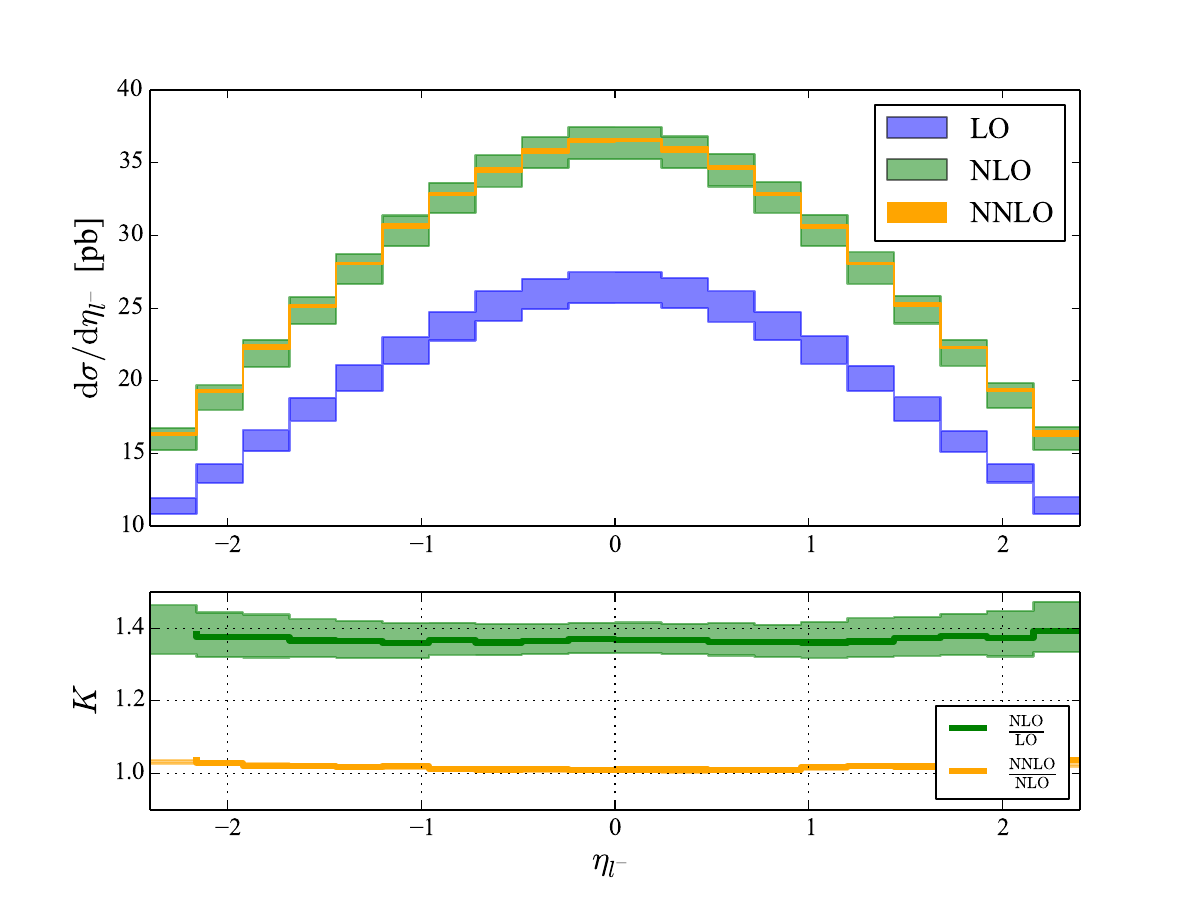}%
\caption{Plot of the lepton rapidity distribution at LO, NLO and NNLO in QCD perturbation theory, for 13 TeV collisions with the central scale $\mu_0=\sqrt{m_{ll}^2+\sum p_{T}^{jet,2}}$.  The $K$-factors are shown in the lower inset.} \label{fig:Yl}
\end{figure}

We now study distributions of the lepton that comes from the $Z \to l^+ l^-$ decay; the anti-lepton distributions are similar.  The lepton transverse momentum distribution at LO, NLO and NNLO in QCD perturbation theory is shown in Fig.~\ref{fig:pTl}.  We focus on the range $p_{T}^{l^-} \leq 180$ GeV due to the small cross section at higher transverse momenta.  There is again a reduction of the scale uncertainty to the percent level when the NNLO corrections are included.  The NNLO corrections rise slightly as $p_{T}^{l^-}$ is increased.  The variation of the $K$-factors that appears for low-$p_{T}^{l^-}$ arises from the leading-order kinematic restriction that $p_{T}^Z>30$ GeV, which occurs because of the $p_{T}^{jet}>30$ GeV cut.  This in turn restricts the allowed values of $p_{T}^{l^-}$ that can occur.  This restriction is lifted at NLO when additional radiation is present, but leads to large corrections near the LO kinematic boundary.  Finally, we show in Fig.~\ref{fig:Yl} the rapidity distribution of the lepton.  The kinematic variation of the $K$-factor is small at both NLO and NNLO, with the corrections being a constant $+40\%$ shift at NLO and nearly zero at NNLO.  Although not shown explicitly here, we find a similar pattern of corrections for the jet and $Z$-boson rapidity distributions.

Before concluding we comment briefly on some computational aspects of our calculation.  It was recently shown that a multi-threaded version of the Vegas integration algorithm~\cite{Lepage:1977sw} could significantly reduce the time needed to obtain NLO cross sections~\cite{Campbell:2015qma}.  We have extended this parallelization to use the MPI protocol in order to allow communication between the separate nodes present on modern computing clusters.  Numerical tests on the  Mira supercomputer at the Argonne Leadership Computing Facility and at the NERSC facility at Berkeley show that our code exhibits strong scaling to the several-thousand node level.  We anticipate that the techniques  we have developed will become increasingly important for theoretical predictions to match the ever-improving quality and precision of high energy collider data.

\section{Conclusions}

In this manuscript we have presented the complete NNLO corrections to the $Z$+jet process in hadronic collisions.  Our calculation utilizes the $N$-jettiness subtraction scheme, which has proven to be a powerful tool for obtaining higher-order QCD cross sections.  We have given phenomenological results for 13 TeV LHC collisions.  The NNLO corrections are small throughout most studied regions of phase space, and are at or below the percent level for $p_T$ values up to 100 GeV.  However, they reach up to 10\% in the tails of the jet and $Z$-boson transverse momentum distributions, and must be included in any comparison of theory with experiment in this region. The corrections to the rapidity distributions of the jet, $Z$-boson and leptons are flat, and are at or below the few-percent level for all scale choices.  The $Z$+jet prediction exhibits an extremely stable perturbative expansion, and upon inclusion of the complete NNLO corrections is ready for a precision comparison with LHC Run II data.
 
The $N$-jettiness subtraction scheme has now been applied to obtain the complete NNLO results for several important LHC processes.  One great virtue of this approach is its simplicity: all complications associated with the double-unresolved singular limit of QCD are handled by the factorization theorem of Eq.~(\ref{eq:fact}).  Another advantage of this approach is the ease with which the necessary numerical integrations can be efficiently run on massively parallel computing platforms. Only the real-radiation integration  in the region $\taun > \tauncut$ is computationally expensive.  The calculational method scales to the largest available computing platforms. The conceptual appeal, simplicity and computational advantages of $N$-jettiness subtraction will make it a powerful tool whenever precision predictions for scattering processes are required.
 
\noindent
\mysection{Acknowledgements}

We thank T.~LeCompte for many helpful discussions.  R.~B. is supported by the DOE contract DE-AC02-06CH11357.  J.~C., K.~E. and W.~G. are supported by the DOE contract DE-AC02-07CH11359.  C.~F. is supported by the NSF grant PHY-1520916.  X.~L. is supported by the DOE grant DE-FG02-93ER-40762.  F.~P. is supported by the DOE grants DE-FG02-91ER40684 and DE-AC02-06CH11357.  This research used resources of the National Energy Research Scientific Computing Center, a DOE Office of Science User Facility supported by the Office of Science of the U.S. Department of Energy under Contract No. DE-AC02-05CH11231.  It also used resources of the Argonne Leadership Computing Facility, which is a DOE Office of Science User Facility supported under Contract DE-AC02-06CH11357.


\begin{thebibliography}{99}

\bibitem{Giele:1993dj} 
  W.~T.~Giele, E.~W.~N.~Glover and D.~A.~Kosower,
  Nucl.\ Phys.\ B {\bf 403}, 633 (1993).

\bibitem{EWcor} 
  A.~Denner, S.~Dittmaier, T.~Kasprzik and A.~Muck,
  JHEP {\bf 1106}, 069 (2011);
W.~Hollik, B.~A.~Kniehl, E.~S.~Scherbakova and O.~L.~Veretin,
  Nucl.\ Phys.\ B {\bf 900}, 576 (2015);
  S.~Kallweit, J.~M.~Lindert, S.~Pozzorini, M.~Schšnherr and P.~Maierhšfer,
  arXiv:1511.08692 [hep-ph].


\bibitem{Ridder:2015dxa} 
  A.~G.~D.~Ridder, T.~Gehrmann, E.~W.~N.~Glover, A.~Huss and T.~A.~Morgan,
  arXiv:1507.02850 [hep-ph].

\bibitem{Aad:2013ysa} 
  G.~Aad {\it et al.} [ATLAS Collaboration],
  JHEP {\bf 1307}, 032 (2013).

\bibitem{Khachatryan:2014zya} 
  V.~Khachatryan {\it et al.} [CMS Collaboration],
  Phys.\ Rev.\ D {\bf 91}, no. 5, 052008 (2015).

\bibitem{radcortalk}
See the talk by T.~Morgan at the 2015 Radcor-Loopfest Symposium,~\url{https://hepconf.physics.ucla.edu/radcor-loopfest/index.html} 

 \bibitem{Boughezal:2015dva} 
  R.~Boughezal, C.~Focke, X.~Liu and F.~Petriello,
  Phys.\ Rev.\ Lett.\  {\bf 115}, no. 6, 062002 (2015).
  
\bibitem{Gaunt:2015pea} 
  J.~Gaunt, M.~Stahlhofen, F.~J.~Tackmann and J.~R.~Walsh,
  arXiv:1505.04794 [hep-ph].

\bibitem{Boughezal:2015aha} 
  R.~Boughezal, C.~Focke, W.~Giele, X.~Liu and F.~Petriello,
  Phys.\ Lett.\ B {\bf 748}, 5 (2015).

\bibitem{Boughezal:2013uia} 
  R.~Boughezal, F.~Caola, K.~Melnikov, F.~Petriello and M.~Schulze,
  JHEP {\bf 1306}, 072 (2013).
  
 \bibitem{Chen:2014gva} 
  X.~Chen, T.~Gehrmann, E.~W.~N.~Glover and M.~Jaquier,
  Phys.\ Lett.\ B {\bf 740}, 147 (2015). 
  
\bibitem{Boughezal:2015dra} 
  R.~Boughezal, F.~Caola, K.~Melnikov, F.~Petriello and M.~Schulze,
  Phys.\ Rev.\ Lett.\  {\bf 115}, no. 8, 082003 (2015).  


\bibitem{Campbell:2010ff} 
  J.~M.~Campbell and R.~K.~Ellis,
  Nucl.\ Phys.\ Proc.\ Suppl.\  {\bf 205-206}, 10 (2010).

\bibitem{wip1}
R.~Boughezal, J.~M.~Campbell, R.~K.~Ellis, C.~Focke, W.~Giele, X.~Liu, F.~Petriello and C.~Williams,
  arXiv:1605.08011 [hep-ph].
  
\bibitem{Stewart:2010tn} 
  I.~W.~Stewart, F.~J.~Tackmann and W.~J.~Waalewijn,
  Phys.\ Rev.\ Lett.\  {\bf 105}, 092002 (2010).

\bibitem{Jouttenus:2011wh} 
  T.~T.~Jouttenus, I.~W.~Stewart, F.~J.~Tackmann and W.~J.~Waalewijn,
  Phys.\ Rev.\ D {\bf 83}, 114030 (2011).

\bibitem{Alioli:2012fc} 
  S.~Alioli, C.~W.~Bauer, C.~J.~Berggren, A.~Hornig, F.~J.~Tackmann, C.~K.~Vermilion, J.~R.~Walsh and S.~Zuberi,
  JHEP {\bf 1309}, 120 (2013);
  S.~Alioli, C.~W.~Bauer, C.~Berggren, F.~J.~Tackmann, J.~R.~Walsh and S.~Zuberi,
  JHEP {\bf 1406}, 089 (2014).

\bibitem{Stewart:2009yx} 
  I.~W.~Stewart, F.~J.~Tackmann and W.~J.~Waalewijn,
  Phys.\ Rev.\ D {\bf 81}, 094035 (2010).

\bibitem{Stewart:2010pd} 
  I.~W.~Stewart, F.~J.~Tackmann and W.~J.~Waalewijn,
  Phys.\ Rev.\ Lett.\  {\bf 106}, 032001 (2011).

\bibitem{Bauer:2000ew} 
  C.~W.~Bauer, S.~Fleming and M.~E.~Luke,
  Phys.\ Rev.\ D {\bf 63}, 014006 (2000);
 C.~W. Bauer, S.~Fleming, D.~Pirjol, and I.~W. Stewart,
\newblock Phys. Rev. {\bf D63}, 114020 (2001), hep-ph/0011336;
  C.~W.~Bauer and I.~W.~Stewart,
  Phys.\ Lett.\ B {\bf 516}, 134 (2001)
  [hep-ph/0107001];
C.~W. Bauer, D.~Pirjol, and I.~W. Stewart,
\newblock Phys. Rev. {\bf D65}, 054022 (2002), hep-ph/0109045;
C.~W. Bauer, S.~Fleming, D.~Pirjol, I.~Z. Rothstein, and I.~W. Stewart,
\newblock Phys. Rev. {\bf D66}, 014017 (2002), hep-ph/0202088.

\bibitem{Gehrmann:2011ab} 
  T.~Gehrmann and L.~Tancredi,
  JHEP {\bf 1202}, 004 (2012).
\bibitem{Gehrmann:2013vga} 
  T.~Gehrmann, L.~Tancredi and E.~Weihs,
  JHEP {\bf 1304}, 101 (2013).

\bibitem{Gaunt:2014xga} 
  J.~R.~Gaunt, M.~Stahlhofen and F.~J.~Tackmann,
  JHEP {\bf 1404}, 113 (2014);
  J.~Gaunt, M.~Stahlhofen and F.~J.~Tackmann,
  JHEP {\bf 1408}, 020 (2014).

\bibitem{Becher:2006qw} 
  T.~Becher and M.~Neubert,
  Phys.\ Lett.\ B {\bf 637}, 251 (2006);
  T.~Becher and G.~Bell,
  Phys.\ Lett.\ B {\bf 695}, 252 (2011).

\bibitem{Boughezal:2015eha} 
  R.~Boughezal, X.~Liu and F.~Petriello,
  Phys.\ Rev.\ D {\bf 91}, no. 9, 094035 (2015).

\bibitem{Becher:2011fc} 
  T.~Becher, C.~Lorentzen and M.~D.~Schwartz,
  Phys.\ Rev.\ Lett.\  {\bf 108}, 012001 (2012).

\bibitem{Dulat:2015mca} 
  S.~Dulat {\it et al.},
  arXiv:1506.07443 [hep-ph].
  
\bibitem{Cacciari:2008gp} 
  M.~Cacciari, G.~P.~Salam and G.~Soyez,
  JHEP {\bf 0804}, 063 (2008).

\bibitem{Lepage:1977sw} 
  G.~P.~Lepage,
  J.\ Comput.\ Phys.\  {\bf 27}, 192 (1978).

\bibitem{Campbell:2015qma} 
  J.~M.~Campbell, R.~K.~Ellis and W.~T.~Giele,
  Eur.\ Phys.\ J.\ C {\bf 75}, no. 6, 246 (2015).

\end{thebibliography}
\end{document}